# Synthesis and characterization of single phase Mn doped ZnO

S. Chattopadhyay[a], S. Dutta[a], A. Banerjee[a], D. Jana[a], S. Bandyopadhyay[a]*, S. Chattopadhyay[b], A. Sarkar[c]

[a]Department of Physics, University of Calcutta, 92 Acharya Prafulla Chandra Road, Kolkata-700 009, West Bengal, India

[b]Department of Physics, Taki Government College, Taki -743429, West Bengal, India

[c]Department of Physics, Bangabasi Morning College, 19 Rajkumar Chakraborty Sarani, Kolkata -700 009, West Bengal, India

---

* Author to whom correspondence should be addressed; electronic mail: sbaphy@caluniv.ac.in

## Abstract

Different samples of $Zn_{1-x}Mn_xO$ series have been prepared by conventional solid-state sintering method. It has been identified, up to what extent of doping enable us to synthesize single-phase polycrystalline Mn doped ZnO samples which is one of the prerequisite for dilute magnetic semiconductor and we have analyzed its certain other physical aspects. In synthesizing the samples proportion of Mn varies from 1 at% to 5 at%. However the milling times have been varied (6, 12, 24, 48 & 96 hours) for only 2 at% Mn doped samples while for other samples (1, 3, 4 & 5 at% Mn doped) the milling time has been kept fixed at 96 hours. Room temperature X-Ray diffraction (XRD) data reveal that all of the prepared samples up to 3 at% of Mn doping exhibit wurtzite-type structure, no segregation of Mn and/or its oxides has been found. The 4 at% Mn doped samples show a weak peak of $ZnMn_2O_4$ apart from usual other peaks of ZnO and the intensity of this impurity peak has been further increased for 5 at% of Mn doping. So beyond 3 at% doping single-phase behavior is destroyed. Band gap for all the 2 at% Mn doped samples have been estimated as between 3.21 to 3.19 eV and reason for this low band gap values has been explained through the grain boundary trapping model. The room temperature resistivity measurement shows increase of resistivity up to 48 hours of milling and with further milling it saturates. The defect state of these samples has been investigated by using positron annihilation lifetime (PAL) spectroscopy technique. Here all the relevant lifetime parameters of positron i.e. free annihilation ($\tau_1$), at defect site ($\tau_2$) and average ($\tau_{av}$) increases with milling time.

**Introduction**

Enormous application of spintronic devices such as spin-valve transistor, spin-light emitting diodes, non-volatile memory, optical isolator, ultra fast optical switches, inspires extensive research work to synthesize Diluted Magnetic Semiconductor (DMS) with III-V and II-VI semiconductors [1-16]. Quite substantial amount of work has been done with transition metal (Tm) doped III-V semiconductor particularly Mn doped GaAs [2,3]. The highest reported ferromagnetic Curie temperature ($T_C$) in this system is ~ 172 K, which is much lower than room temperature. Dietl et al. [17] predicted high temperature ferromagnetism in magnetically-doped wide band-gap p-type semiconductors particularly in ZnO, GaN, GaAs and ZnTe. This fact has motivated researchers to study the properties of Tm doped semiconductors. Room temperature ferromagnetism has been reported by Sharma et al. [1], Milivojevic et al. [10] in polycrystalline $Zn_{1-x}Mn_xO$ samples, prepared from solid state sintering route. ZnO based DMS system have some advantages over the other because of its some unique characteristics like large band gap (~3.4 eV), large exciton binding energy at room temperature (~ 60 meV), high optical gain (300 $cm^{-1}$) very short luminescence life time [11] which are required for various optoelectronic [12] and magnetoptical [13] devices.

In this work, we have made a systematic study of structural, optical, electrical properties and defect analysis through Positron annihilation lifetime (PAL) spectroscopy of 5 different 2 at% Mn doped samples prepared by solid state sintering method at different milling time. Study of structural properties has also been performed for 1, 2, 3, 4 and 5 at% Mn doped samples prepared with 96 hours of milling. Structural characterization was made through X ray diffractometer (XRD). A primary requisition for DMS material is that structure of the material must be of single phase. This is required because the chances of ferromagnetic ordering due to segregation of magnetic ions arising out of doping of transition metals into an impurity phase or any other precipitates would be zero. Though it is true that some of the impurity phase(s) in ZnMnO bear ferromagnetism; but ferromagnetism originating from the impurity phases cannot have the desired properties for spintronic devices such as spin injection [18]. So the extent of doping by Tm has definitely a limit beyond which multi-phase components starts to arise. In this work we have estimated the maximum amount of dopant (Mn) incorporation in the host material ZnO for samples derived out of solid state sintering method so that no signature of impurity phase has been developed.

Room temperature resistivities were also recorded and optical studies were done by UV-Visible (UV-Vis) absorption spectroscopy. It is an established fact that ferromagnetism in DMS materials is carrier-mediated [19]. In Mn doped ZnO the study on defect states will be crucial as defects are proposed to generate free carriers [20] in ZnO. Such defects may play a significant role in achieving ferromagnetism [21]. In his pioneering work, Dietl pointed out that, defect state plays an important role for showing ferromagnetism in these types of materials [17]. Iusan et al. [22] and Hasu et al. [23] have done interesting theoretical work on establishing the role of defects on ferromagnetism in Tm-doped ZnO. But there is hardly any experimental study to analyze the role of defect state not only in Tm doped ZnO samples but also for any other DMS materials. We have made a systematic study of the $Zn_{0.98}Mn_{0.02}O$ samples using room temperature PAL. Positron lifetimes inside a material (~ 100-400 ps) can be measured from PAL spectroscopy. Positrons from a radioactive source (here $^{22}Na$) are injected into the material and get thermalized within 1-10 ps. Then they annihilate with an electron of that material. It is well known that positrons are preferentially populated in low electron concentration regions. Positrons can be trapped in defects present in the material. The lifetime of trapped positrons in defects is comparatively longer with respect to those annihilate at defect free regions. So, the analysis of PAL spectrum may provide some interesting results regarding the nature and abundance of defects.

**Experimental details**

Five different 2 at% Mn doped ZnO ($Zn_{1-x}Mn_xO$) have been prepared through conventional solid-state sintering method [1] using ZnO (99.99%, Sigma-Aldrich, Germany) and $MnO_2$ (99.9%, Sigma-Aldrich, Germany) at different milling time (6, 12, 24, 48 and 96 hours). The stoichiometric amounts of samples have been taken. The samples were milled first with different duration by using "Fritsch planetary mono mill" (Model No: pulverisette 6). The milling has been performed in two stages. The first milling time was taken one third of the total milling time for a particular sample. Next they were sintered first at 400 $^o$C for 8 hours and finally, after making pallets, they were sintered at 500 $^o$C for 12 hour using digital furnace. In between the two sintering process the samples were milled with remaining two third of the total milling time. The ball to mass ratio was maintained at 1:1 throughout the milling process. The final annealing temperature was 500 $^o$C as it was reported that the samples annealed at 500 $^o$C shows room temperature ferromagnetism [1,18,24] and decreases with annealing above 500 $^o$C [1,24]. Garcia et al. further showed that in presence of Zn, the transformation of $Mn^{+4}$ states to $Mn^{+3}$ states starts about 200 $^o$C. The transformation is completed at about 600 $^o$C. Now ferromagnetism in MnZnO system arises when $Mn^{+3}$ and $Mn^{+4}$ states coexist. So we expect a substantial portion of $Mn^{+4}$ states transformed into $Mn^{+3}$ states at the annealing temperature around 500 °C. It is the reason for choosing 500 °C as final annealing temperature. We have also prepared 5 different samples of $Zn_{1-x}Mn_xO$ series by varying the proportion of Mn from 1 to 5 at% with 96 hours of milling keeping all other conditions unaltered.

The XRD data have been collected in a Philips PW 1830 automatic powder diffractometer with Cu $K_α$ radiation. The range of scanning is 10°- 80° (2θ) in a step size 0.01°. We have calculated the crystallite-size using Scherrer's formula [25]. The contribution of instrumental broadening has been taken into account.

The variation of band gap with milling time has been investigated by UV-Vis absorption spectroscopy using Hitachi U-3501 spectrophotometer in the wavelength range of 200–1600 nm at room temperature.

Room temperature resistivity measurement has been carried out in usual two-probe technique using Keithley 6514 Electrometer, because of high resistive samples.

For positron annihilation study, a 10-μCi $^{22}$Na positron source (enclosed in thin mylar foil) has been sandwiched between two identical plane faced pallet of the samples. We have measured the PAL spectra with a fast-slow coincidence assembly [26] having 182 ± 1 ps time resolution. Measured spectra have been analyzed by computer program PATFIT-88 [27] with necessary source correction to obtain the possible lifetime components $τ_i$, and their corresponding intensities $I_i$.

**Result and discussion**

XRD patterns of all $Zn_{0.98}Mn_{0.02}O$ (prepared at different milling time, 6, 12, 24, 48 and 96 hours) are shown in figure 1. The indices in the spectra of figure 1 indicate the expected positions of the peaks for the wurtzite crystal structure of ZnO, no signature of impurity peaks have been observed [10]. This indicates that Mn might have been substitutionally incorporated in ZnO lattice and so no impurity phase has been formed. The above observation is also true for the samples doped with 1 and 3 at% Mn prepared at 96 hours of milling. When the proportion of Mn has been increased to 4 at% a weak (112) peak of $ZnMn_2O_4$ has been observed at 29.11º (2θ), with further increase of proportion of Mn to 5 at% the intensity of this (112) peak of $ZnMn_2O_4$ has been slightly increased, as shown in figure 2. The growing tendency of intensity of the impurity peak in cases of 5 at% doped sample has been shown clearly in figure 3 with exaggerated plotting only in the range of 2θ from 28º to 30º for 3, 4 and 5 at% Mn doped samples. From figure 2 it may appears that the growing up of (112) peak of $ZnMn_2O_4$ around 29º starts from 3 at% Mn doped samples but careful observation of figure 3 strongly shows that the (112) peak of $ZnMn_2O_4$ starts growing from 4 at% doping. Actually the formation of impurity phase in case of lightly doped semiconductors depends on inability of the power of the dopants (here transition metal Mn) to replace substitutionally the cation (here Zn) of the host compound semiconductor. Now, up to 3 at% level of doping Mn can substitutionally replace Zn but when the degree of doping exceeds further to 4 at% and above not all of the Mn atoms replace Zn atoms rather segregated as $ZnMn_2O_4$. As single phase formation is one of the primary criterions in our samples, so in this solid state sintering method for Mn doped ZnO samples doping concentration can not be increased beyond 3 at%. The effect of crystallite-size-induced broadening has been analyzed using Sherrerr's formula using the (101) peak, as shown in the inset of figure 1. The crystalline size can be obtained, using the following relation:

$$D = \frac{K\lambda}{\beta \cos\theta}$$

where *β* is FWHM in radians, *D* is the average crystallite size, *K* is the shape factor (usually taken 0.89), *λ* is the x-ray wavelength and *θ* is the Bragg angle. The estimated grain size of the 2 at% Mn doped samples has been found to be varying from 34.5 nm to 30 nm and plotted in the inset of figure 1. The values of grain size for different 2 at% Mn doped samples prepared at different milling time are shown in table 1. the little variation of grain size data indicates that milling time has no direct correlation with grain size.

UV-Vis absorption spectra of all the 2 at% Mn doped samples are shown in the inset of figure 4. According to Tauc et al. [28] and Pancov [29] for a given transition, photon energy $h\nu$ can be related to band gap energy ($E_g$) by the following expression:

$$\alpha = \frac{A(h\nu - E_g)^{\frac{m}{2}}}{h\nu}$$

where m = 1 for a direct transition and m = 4 for an indirect transition, *α* is the absorption coefficient and *A* is a constant. As ZnO is a direct band gap semiconductor we have calculated the band gap of these sample from linear fitting to the *(αhν)²* against energy (*E*) plot as shown in figure 4 and the value of band gap obtained for all the samples is in between 3.21 to 3.19 eV which is slightly lower in generally reported [30] value as 3.4 eV. In one of the earlier work [31] of our group it was calculated as 3.22 eV for pure ZnO powders. Nevertheless, the reduction in band gap might be due to a smaller average grain size and possibly lower carrier concentration (because of the highvalue of resistivity to be shown bellow) in our samples. To explain this small variation of the band gap of our Mn-doped ZnO samples, we introduce a variant of a simple model proposed earlier for ZnO film on fused silica [32] and glass substrate [33]. As a result of this intuitive picture, periodic variations in potential within the grain do occur due to trapping of impurities arising out of Mn doping. This manifests naturally to a non-uniform variation of the electric fields at the grain boundaries. This causes a vertical band gap, which is the difference between the highest point of valence band and lowest point of conduction band. An idealized periodic potential diagram with the important parameters marked on it is shown schematically in figure 5. From the diagram it can be seen that the effective band gap $E_g^{eff}$ is given by:

$$E_g^{eff} = E_g^0 - \phi + \xi_1$$

where $E_g^0$ is the original gap, $\phi$ the barrier height at the grain boundary and $\xi_1$ the first quantized level of a 1D gas. Using Scherrer's formula, we find the variation of average grain size lies between 30-34 nm. This indicates a band bending of 0.2016 eV. This is consistent with the observed difference of the band gap of the samples. This justifies that the model chosen for analyzing the optical data seems to be proper one.

In figure 6, the variation of room temperature resistivity with milling time has been shown. The room temperature resistivity increases with milling time up to 48 hours of milling and later there is an indication of attaining saturation in resistivity value, as is also evident from table.1. Moreover, the value of the resistivity is lower for the 6 and 12 hours milled samples compared to that of pure ZnO (8.4 x $10^7$ Ω-cm) [33]. But for samples milled with longer periods such as 24, 48 and 96 hours the resistivity becomes very high. The possible reason might be behind this variation that with longer period of milling the inherent non-stoichiometry present in the ZnO matrix might be dissolved. The oxygen vacancy present in the system of ZnO is responsible for comparatively low value of resistivity in case of the sample with milling time 6 and 12 hours. There are two competitive effects: sintering at 500 $^oC$ increases the density of oxygen vacancy and milling tries to make the sample stoichiometric by decreasing the density of oxygen vacancy. When the milling time is only 6 or 12 hours the former one predominates over the later one so the resistivity decreases and reaches below the value of pure ZnO. The resistivity of pure ZnO was observed without sintering and milling also. But when the period of milling gradually increases from 12 hours to 48 hours and as period and temperature of sintering remains the same, the later effect dominates over the former one giving rise to a significant increase in resistivity. With further increase in milling time, the specimen becomes less oxygen deficient, its resistivity increases and nearly saturates after 48 hours of milling.

To elucidate the above intuitive picture, we have performed PAL analysis and the results are shown in figure 7 and table 2. All lifetime spectra are best fitted with three components lifetime fit. The longest ($\geq$ 1248 ps with intensity 3% to 4%) among them is the third one ($\tau_3$). Positron annihilation from orthopositronium like atoms contributes for $\tau_3$. Orthopositronium formation inside the micro voids [24] is always present within the materials. It decays into parapositronium through pickoff annihilation, giving rise to such a large lifetime. In polycrystalline samples voids are always present that favors positronium formation [31,34]. But positronium is normally found in a defect free solid in a self trapped state, i.e. positronium creates by itself a 'cage' by pushing away the surrounding atoms [35]. So, the origin of $\tau_3$ within materials is a different process and not related to the positron trapping at defects. The shortest lifetime component is $\tau_1$, comes due to the free annihilation of positrons [26, 34,36]. The effects of small vacancies (like mono-vacancies etc.) [37-40] or shallow vacancies (like oxygen vacancies [39] in ZnO) may also be related with $\tau_1$ in case of disordered systems. Though $\tau_1$ itself is a weighted average of free and trapped annihilation but these sites are not major positron traps. We have found (shown in table 2) that there is an increase in the value of $\tau_1$ with milling time. The above mentioned two competitive effects, sintering and milling, may also be responsible for this kind of variation of $\tau_1$. Oxygen vacancy have been increased by sintering at 500 $^oC$ which in tern increases the electron density within the sample whereas milling tries to make the sample stoichiometric by decreasing the density of oxygen vacancy that imply a decrease in free electron density. We have gradually increased the period of milling from 6 hours to 96 hours with the period and temperature of sintering remains same. With the increase of the milling time the effect of milling predominates over the sintering effect so the available electron density reduced and it is manifested both in resistivity data and $\tau_1$. Here a point is noteworthy that reduction in carrier density may originate due to the localization of electrons into some open volume defects which in tern increase both resistivity and $\tau_1$.

The most important component of positron lifetime is $\tau_2$ that indicates qualitatively the nature and size of the vacancy [39] and its relative intensity gives a quantitative measure of abundance of that vacancy with respect to some standard of the same sample. We have found considerable increase in the $\tau_2$ values with increasing milling time from 6 hours to 96 hours (figure 7(a)), the variation of corresponding intensities ($I_2$) is given in the inset of figure 7(a). From $\tau_1$, $I_1$, $\tau_2$ and $I_2$ the average positron lifetime $\tau_{av} \left[ = \dfrac{\tau_1 I_1 + \tau_2 I_2}{I_1 + I_2} \right]$ can be constructed. $\tau_{av}$ shows an overall increase with milling time similar to that of $\tau_2$. This might occur because of the increase in defect clustering near the grin boundary due to mechanical milling. The thermal energy given to the system by sintering process may help the intra-grain Zn vacancies to approach towards the grain surfaces, which are the universal sink of defects. In this way, small size Zn vacancies (mono-vacancies, etc.) assemble near the grain surfaces give rise to $\tau_1$ (shown in table 2). A fraction of such mono-vacancies joins together to form larger size vacancy clusters causing an increase in the value of $\tau_2$ also.

**Conclusion**

In this present work we have characterize the $Zn_{1-x}Mn_xO$ using room temperature XRD, UV-Vis absorption spectroscopy, room temperature resistivity and room temperature PAL spectroscopy. From the results of these characterization the following conclusions can be drown.

Room temperature XRD results indicate that all the samples, except 4 and 5 at% Mn doped sample exhibit wurtzite-type structure, similar to that of stoichiometric ZnO, no segregation of Mn and/or its oxides has been found. This signifies that, Mn might have been substitutionally incorporated in the ZnO lattice for these samples. The 4 and 5 at% Mn doped sample shows a weak peak of $ZnMn_2O_4$ apart from usual other peaks of ZnO. The intensity of this peak increases with increasing Mn concentration. It is highly indicative in this purpose regarding the extent of doping of magnetic species, which will be useful to achieve intrinsic ferromagnetic ordering.

Band gap for all the Mn doped samples have been estimated as in between 3.21 to 3.19 eV. This indicate that slight decrement in band gap values for Mn doped ZnO samples in comparison to un-doped ZnO sample is due to band bending. The reason for this bending is shown to be the periodic variation of grain boundary potential. The amount of band bending estimated from grain size (calculated from XRD data) is consistent with the values of the band gap calculated from optical measurements.

The room temperature resistivity increases from 4.64 to 239.14 MΩ cm with milling time variation from 6 to 48 hours and for the sample milled for 96 hours, there is an indication of saturation in the resistivity values. The possible reason might be behind this that with longer period of milling (from 6 to 96 hours) the reduction of proportion of the oxygen vacancies, which are one of the sources of free electrons in the ZnO matrix, is reduced.

PAL analysis shows that there is an increase in the value of $\tau_1$ with milling time. This is due to decrease of available electron density. Electrons may localize into some open volume defects, which in tern increase both resistivity and $\tau_1$. The considerable increase in the $\tau_2$ and $\tau_{av}$ values with milling time has been found and it is due to the increase in defect clustering near the grin boundary due to mechanical milling.

**Acknowledgements**

We are grateful to DST-FIST for providing financial assistance. Two of the authors (SC[a] and SD) are grateful to Government of West Bengal for providing financial assistance in form of University Research Fellowship. One of the authors (SB) also thankful to DST for providing grant in a research project with sanction no. SF/FTP/PS-31/2006 dated 20.09.2007. We are also like to acknowledge Mr. Manas Sutradhar, Department of Chemistry, University of Calcutta for helping us in performing UV-Vis spectroscopy experiment.

**Figure Caption**

Figure 1 XRD Spectra of (a) 6, (b) 12, (c) 24, (d) 48 and (e) 96 hours milled 2 at% Mn doped ZnO samples. Inset variation of grain size with milling time of 2 at% Mn doped ZnO samples prepared at different milling times as calculated using Sherrerr formula using (101) peak.

Figure 2 XRD Spectra 96 hour milled (a) 1 at%, (b) 2 at%, (c) 3 at%, (d) 4 at%, (e) 5 at% Mn doped ZnO samples.

Figure 3 exaggerated plotting of XRD spectra in the range of 2θ from 28º to 30º for 3, 4 and 5 at% Mn doped ZnO samples.

Figure 4 Energy Vs. $(\alpha E)^2$ plot of the 2 at% Mn doped ZnO samples prepared by varying milling time. Inset UV-VIS spectra of the 2 at% Mn doped ZnO samples prepared by varying milling times.

Figure 5 Idealized 1D potential variation for several grains which are assumed to be the same size as well as the same trap density

Figure 6 Variation of room temperature resistivity with milling time for different 2 at% Mn doped ZnO samples.

Figure 7 Variation of (a) positron lifetime annihilating at defect site ($\tau_2$) (b) average positron lifetime ($\tau_{av}$) with milling time for 2 at% Mn doped ZnO samples. The inset shows variation of $I_2$ (Intensity corresponding to $\tau_2$) with milling time.

| Milling Time (hours) | Grain Size (nm) | Band gap (eV) | Room Temperature Resistivity (MΩ-cm) |
|---|---|---|---|
| 6 | 34.50 ± 0.24 | 3.21 | 4.64 ± 0.12 |
| 12 | 34.49 ± 0.27 | 3.21 | 72.09 ± 0.57 |
| 24 | 32.44 ± 0.18 | 3.20 | 219.44 ± 5.59 |
| 48 | 31.29 ± 0.20 | 3.19 | 250.58 ± 6.39 |
| 96 | 29.99 ± 0.22 | 3.19 | 239.19 ± 5.65 |

Table. 1 Value of grain size, Band gap and room temperature resistivity of $Zn_{0.98}Mn_{.02}O$ samples prepared at different milling time

| Milling Time (hours) | $\tau_1$ (ps) | $I_1$ (%) | $\tau_3$ (ps) | $I_3$ (%) |
|---|---|---|---|---|
| 6 | 171 ± 1 | 30.56 ± 0.1 | 1349 ± 31 | 3.6 ± 0.2 |
| 12 | 195 ± 1 | 52.21 ± 0.1 | | |
| 24 | 163 ± 1 | 22.30 ± 0.1 | 1317 ± 32 | 3.7 ± 0.2 |
| 48 | 190 ± 1 | 40.23 ± 0.1 | 1248 ± 31 | 3.9 ± 0.2 |
| 96 | 222 ± 1 | 59.30 ± 0.1 | 1406 ± 48 | 3.2 ± 0.2 |

Table. 2 The fitting parameters found from positron annihilation lifetime measurement on $Zn_{0.98}Mn_{.02}O$ samples prepared at different milling time

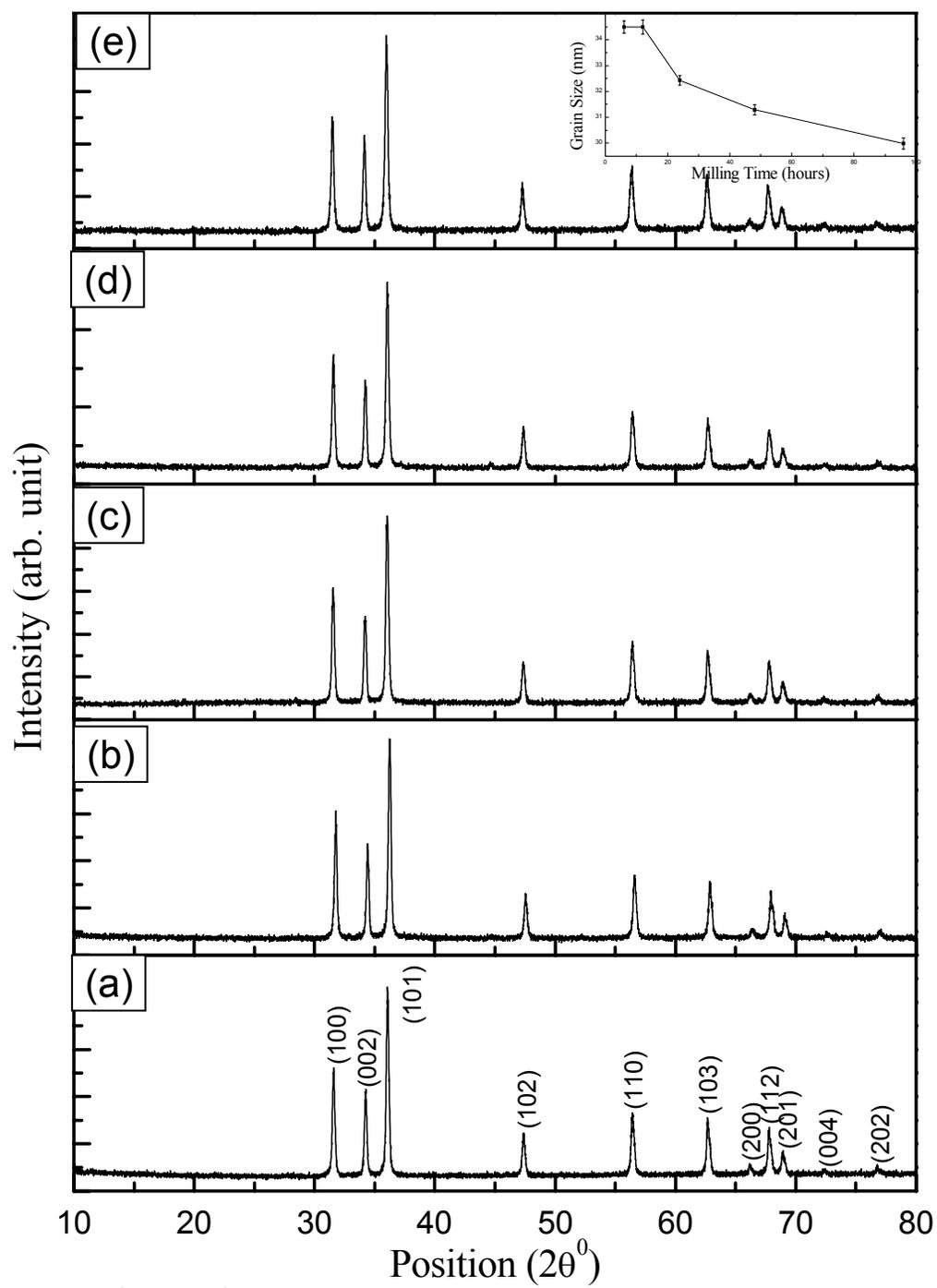

**Figure 1**

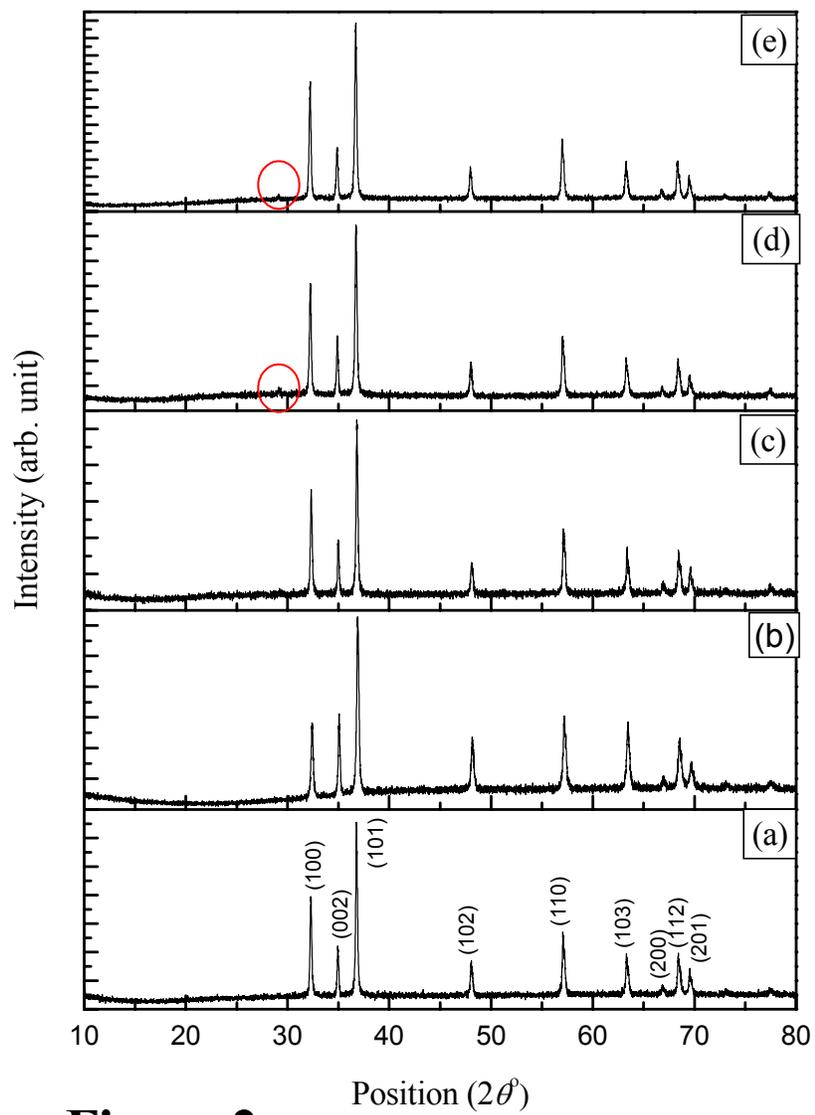

**Figure 2**

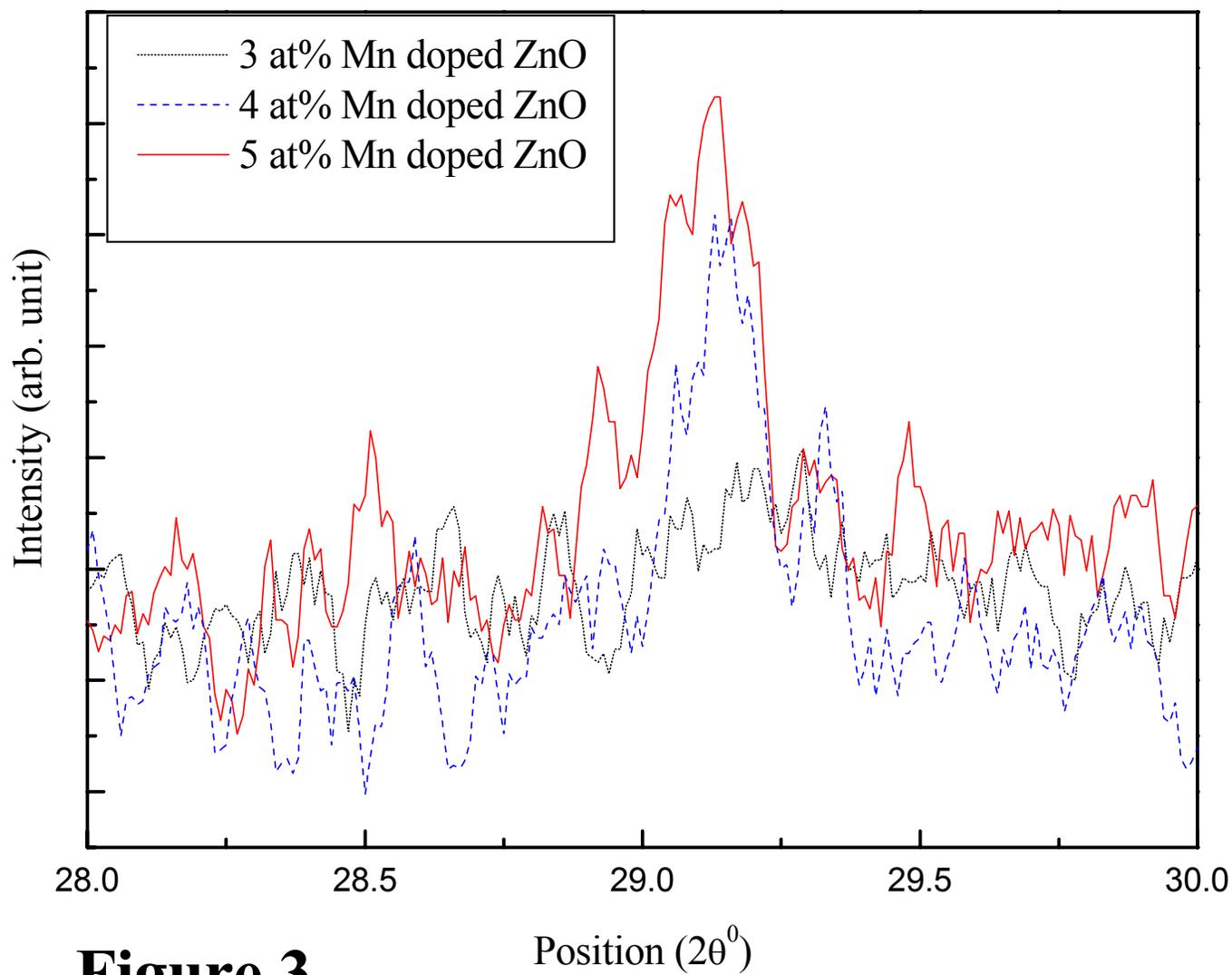

**Figure 3**

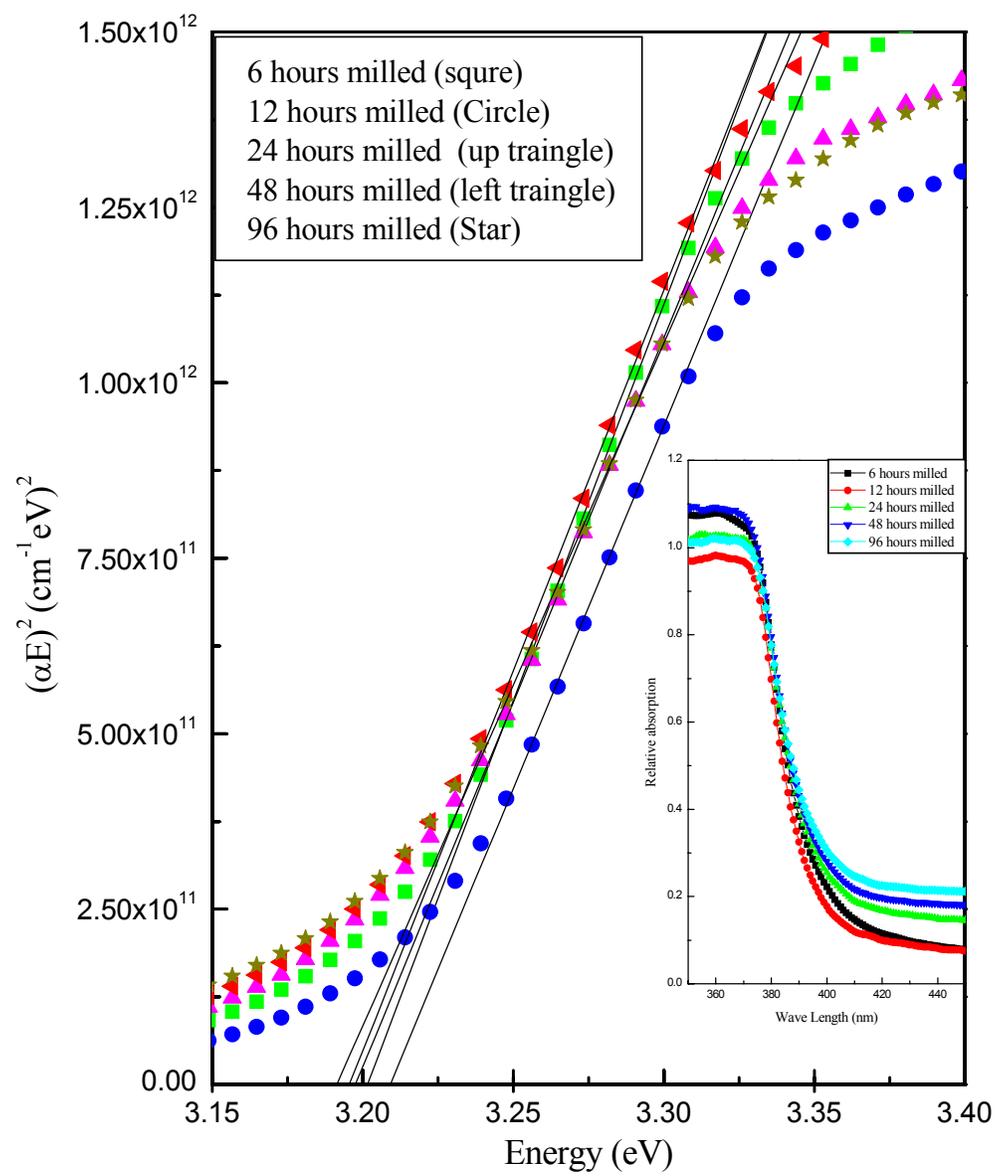

**Figure 4**

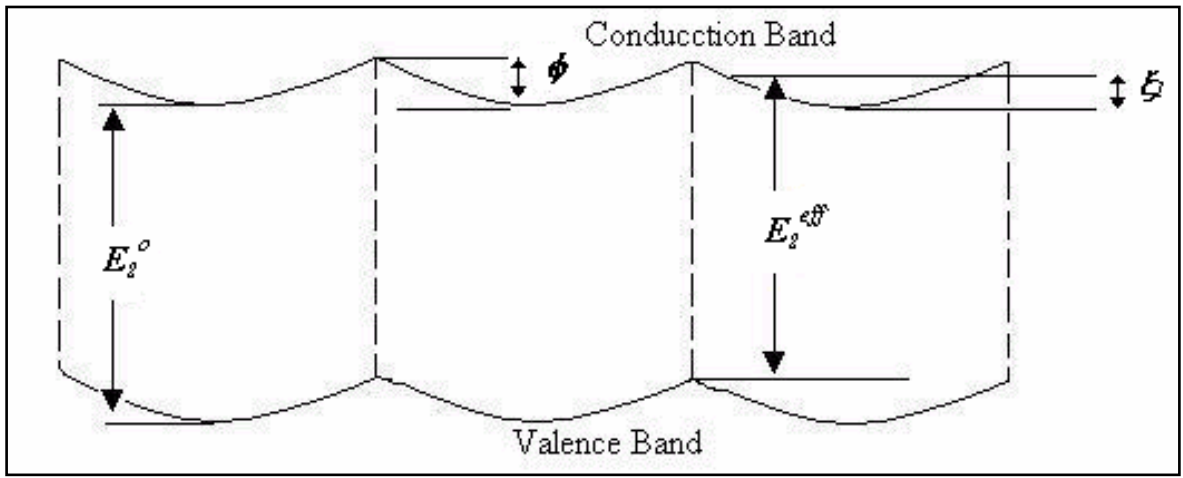

**Figure 5**

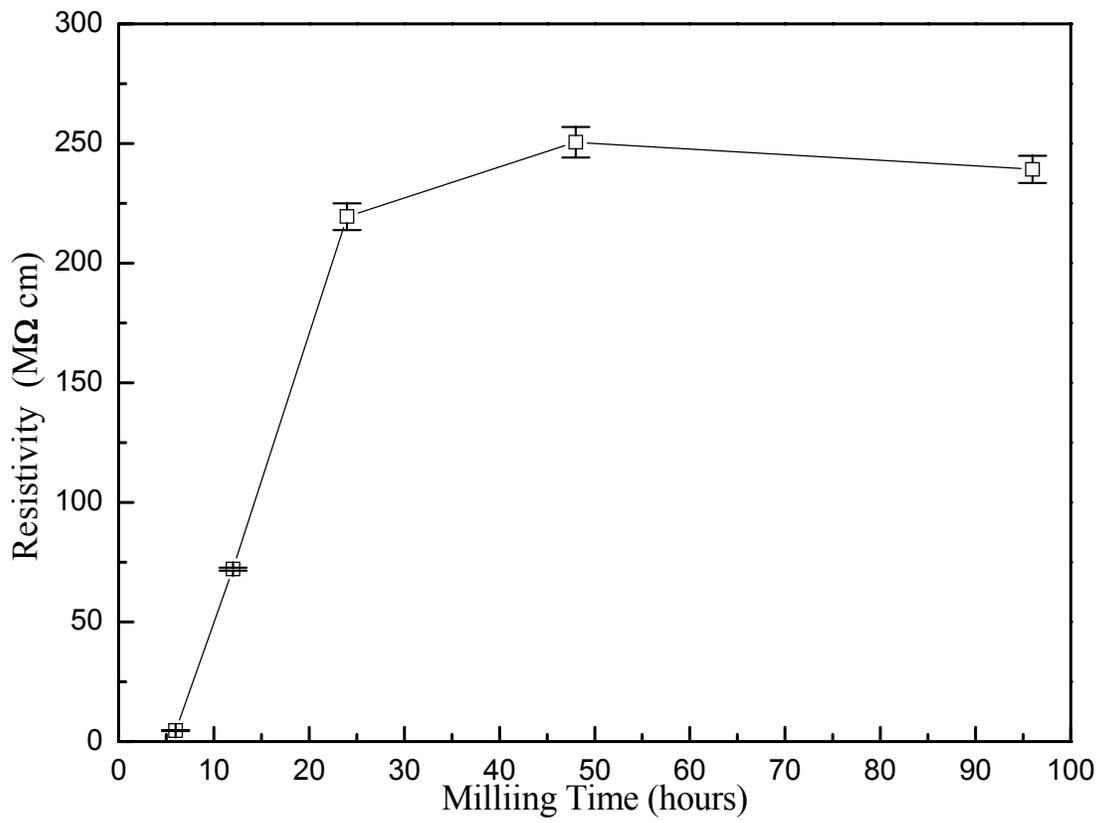

**Figure 6**

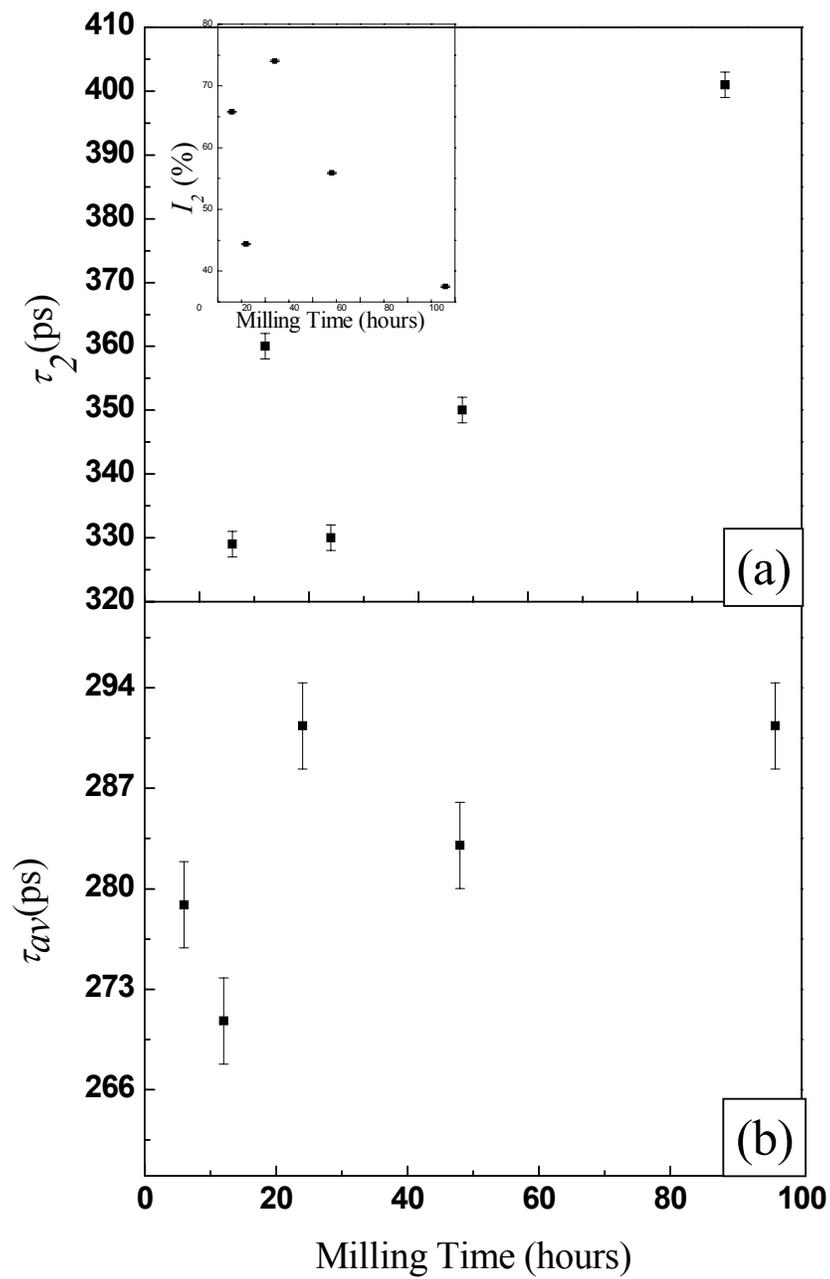

**Figure 7**